\documentclass[pre,aps,showpacs,twocolumn,superscriptaddress]{revtex4}

\usepackage{graphicx}
\usepackage{bm}
\usepackage{amsfonts}
\usepackage{color}

\usepackage{amsmath}    
\usepackage{epsfig}
\usepackage{subfigure}  
\usepackage{hyperref}   
\usepackage{bm}
\usepackage{amssymb}

\newcommand{\romeaddress}{Department of Physics and INFN, University of
Rome ``Tor Vergata'',
Via della Ricerca Scientifica 1, 00133, Rome, Italy.}

\newcommand{\ictsaddress}{International Centre for
  Theoretical Sciences, Tata Institute of Fundamental Research,
  Bangalore 560089, India.}
\newcommand{\iiscaddress}{Centre for Condensed Matter Theory,
Department of Physics, Indian Institute of Science, Bangalore 560012, India.}
\def\la{\left\langle}
\def\ra{\right\rangle}

\def \bfx {{\bf x}}
\def \bfr {{\bf r}}
\def \bfu {{\bf u}}
\def \bff {{\bm f}}
\def \bfb {{\bf b}}

\begin{document}

\noindent\title{Dynamic multiscaling in magnetohydrodynamic turbulence\footnote{Postprint version of the manuscript published in Phys. Rev. E 
{\bf 94}, 053101 (2016)}}
\author{Samriddhi Sankar Ray}
\email{samriddhisankarray@gmail.com}
\affiliation{\ictsaddress}
\author{Ganapati Sahoo}
\email{ganapati.sahoo@gmail.com}
\affiliation{\romeaddress}
\author{Rahul Pandit}
\email{rahul@physics.iisc.ernet.in}
\affiliation{\iiscaddress}
\begin{abstract}

We present the first study of the multiscaling of time-dependent velocity and
magnetic-field structure functions in homogeneous, isotropic
magnetohydrodynamic (MHD) turbulence in three dimensions. We generalize the
formalism that has been developed for analogous studies of time-dependent
structure functions in fluid turbulence to MHD. By carrying out detailed
numerical studies of such time-dependent structure functions in a shell model
for three-dimensional MHD turbulence, we obtain both equal-time and
dynamic scaling exponents. 

\end{abstract} 
\maketitle 

\section{Introduction \label{intro}}

Velocity and magnetic-field structure functions, which are
moments of the probability distribution function (PDF) of the
differences of fluid velocities and magnetic fields between two
points separated by a distance $r$ (a precise definition is given
below), are often used to characterize the statistical properties
of homogeneous, isotropic magnetohydrodynamic (MHD)
turbulence~\cite{sahoomhd,basu98}.  If $r$ is in the {\it
inertial range} of scales that lie between the large length scale
$L$, at which energy is pumped into the system, and the fluid and
magnetic dissipation scales $\eta_d^u$ and $\eta_d^b$,
respectively, at which dissipation becomes significant, these
structure functions scale as a power of $r$. This is similar to
the power-law behaviors of correlation functions at a critical
point in, say, a spin system, as has been elucidated for {\it
equal-time} structure functions in many papers: It turns out that
the simple scaling we see at most critical points must be
generalized to multiscaling in turbulence; i.e., an infinity of
exponents is required to characterize the inertial-range
behaviors of structure functions of different orders as discussed, e.g.,
for fluid turbulence in Ref.~\cite{book-frisch} and for
three-dimensional (3D) MHD turbulence in Ref.~\cite{sahoomhd}.

The scaling behaviors of correlation functions at a critical
point~\cite{book-chaikin} arise from the divergence of a
correlation length $\bar{\xi}$ at the critical point: in a simple
spin system $\bar{\xi} \sim \bar{t}^{-\nu}$ if the magnetic field
$H = 0$, where the reduced temperature $\bar t\equiv (T -
T_c)/T_c$, $T$ is the temperature, $T_c$ the critical
temperature,  and $\nu$ is an equal-time critical exponent that
is universal (in a given universality class). In the vicinity of
a critical point the relaxation time $\tau$, which follows from
{\it time-dependent} correlation functions, scales, according to
the {\it dynamic-scaling Ansatz}, as 
\begin{equation} 
\tau \sim
\bar{\xi}^{z}; 
\label{eq-ch5:ds} 
\end{equation} 
$z$ is known as the dynamic-scaling exponent. There has been considerable
progress
~\cite{lvov97,mitra04,mitra05,belinicher87,kaneda99,hayot98,hayot00,pandit08,ray08,rayprl}
in developing the analog of such a dynamic-scaling {\it Ansatz} for {\it
time-dependent} structure functions in homogeneous, isotropic fluid turbulence.
In this paper we present the first study of time-dependent structure functions
in MHD turbulence. We first set up the formalism in the framework of the 3D MHD
equations and then carry out explicit numerical studies of a shell model for 3D
MHD turbulence ~\cite{basu98}.  We show that (a) an infinity of
dynamic-multiscaling exponents is required and (b) that these exponents depend
on the precise way in which times, e.g., the integral-time scales (defined
below), are extracted from time-dependent structure functions. 

Since the work of Obukhov~\cite{obukhov}, Desnyansky and
Novikov~\cite{desnovikov}, and Gledzer, and Ohkitani and
Yamada~\cite{goy1,goy2} (GOY), shell models have been used to study the
multiscaling of velocity structure functions in fluid
turbulence~\cite{book-frisch,jensen91,pisarenko93,DharPraman97,DharPRL97,Biferale03,bohrbook,ditlevsenbook,sabrashell}.  Generalizations of these
models have been used to study magnetohydrodynamic (MHD)
turbulence~\cite{basu98,mhdshell2,mhdshell3,moremhdshell1,moremhdshell2,
moremhdshell3}, Hall-MHD
turbulence~\cite{hallmhdshell1,hallmhdshell2,hallmhdshell3, hallmhdshell4}, the
effects of polymer additives on fluid turbulence~\cite{polymershell},
two-dimensional fluid turbulence~\cite{2Dshellmodel} and turbulence in between
two and three dimensions~\cite{2D3Dshellmodel}, turbulence in rotating
systems~\cite{rotatingturbulence} and in binary-fluid
mixtures~\cite{symmbinfluid}, superfluid
turbulence~\cite{wacks2011shellmodel,procacciashell3He,procacciaintermittencyshellmodel,vishwanathshellmodel},
and the dynamic multiscaling of time-dependent structure functions in
fluid~\cite{mitra04,pandit08,ray08} and
passive-scalar~\cite{mitra05} turbulence. It behooves us, therefore, to
examine first the dynamic multiscaling of structure functions in a shell model
for three-dimensional MHD (3D MHD). 

In fluid and passive-scalar turbulence, dynamic-multiscaling exponents are
related by {\it linear bridge relations} to equal-time multiscaling
exponents~\cite{pandit08,ray08,rayprl,mitra05}; we have not been able to
find such relations for MHD turbulence so far.  Therefore,  we obtain
equal-time and time-dependent structure functions for a shell model for 3D MHD
turbulence and, from these, equal-time and dynamic multiscaling exponents. We
then try to see if these suggest any bridge relations.

The remaining part of this paper is organized as follows:
Section~\ref{sec:eqsf} contains an introduction to the 3D MHD equations and the
multiscaling of equal-time magnetic and fluid structure functions.
Section~\ref{sec:dm} is devoted to a discussion of the dynamic multiscaling of
time-dependent structure functions in fluid turbulence, in general, and 3D MHD
turbulence, in particular. In Section~\ref{sec:shell} we describe a shell model
for 3D MHD turbulence and give the details of our numerical studies of this
model.  Section~\ref{sec:results} presents results from our simulations for
time-dependent structure functions; and in Section~\ref{sec:concl} we make some
concluding remarks.

\section{Magnetohydrodynamics and Multiscaling\label{sec:eqsf}} 

The flow of conducting fluids can be described by the following 3D MHD
equations~\cite{MHD-reviews} for the velocity field $\bfu(\bfx,t)$ and the
magnetic field $\bfb(\bfx,t)$ at point $\bfx$ and time $t$: 
\begin{eqnarray}
\frac{\partial\bfu}{\partial t}+(\bfu\cdot\nabla)\bfu &=&
\nu \nabla^2\bfu-\nabla\bar{p}+\frac{1}{4\pi}(\bfb\cdot\nabla)
\bfb+\bff_u; \label{eq-ch5:vel}\\
\frac{\partial\bfb}{\partial t}&=&\nabla\times(\bfu\times\bfb)
+\eta\nabla^2\bfb + \bff_b; \label{eq-ch5:mag}
\end{eqnarray}
here $\nu$ and $\eta$ are the kinematic viscosity and the magnetic diffusivity,
respectively, and the effective pressure $\bar{p}=p+(b^2/8\pi)$, where $p$ is
the pressure; $\bfu$ and $\bfb$ have the same dimensions in this
formalism~\cite{MHD-reviews}.  For low-Mach-number flows, to which we restrict
ourselves, we use the incompressibility condition $\nabla\cdot\bfu = 0$, which
we use, along with the constraint $\nabla \cdot\bfb = 0$,  to eliminate the
pressure $\bar{p}$ in Eq. (\ref{eq-ch5:vel}). We choose a uniform density
$\rho=1$ and the external forces $\bff_u$ and $\bff_b$ inject energy into the
conducting fluid, typically at large length scales.  In decaying turbulence,
the external forces are absent.  The dimensionless parameters that characterize
3D MHD turbulence are the kinetic Reynolds number ${\rm Re} \equiv (\ell
v)/\nu$, the magnetic Reynolds number ${\rm Re_M} \equiv (\ell v)/\eta$, and
the magnetic Prandtl number ${\rm Pr_M} \equiv {\rm Re_M/Re}=\nu/\eta$; here
$\ell$ and $v$ are the characteristic length and velocity scales of the flow.

\begin{table*}
{\begin{tabular}{|c|c|c|c|c|c|c|c|c|c|}
\hline
${\rm Pr_M}$ & $\nu$ & $\eta$ & $u_{rms}$ & $b_{rms}$ & $\lambda$ & ${\rm Re}_{\lambda,u}$ & ${\rm Re}_{\lambda,b}$ & $\tau_u$ & $\tau_b$\\
\hline
0.1 & $10^{-7}$ & $10^{-6}$ & 0.84 & 1.27 & 0.879 & 7425500 & 742550 & 7.24 & 15.70\\
\hline
1.0 & $5 \times 10^{-7}$ & $5 \times 10^{-7}$ & 0.85 & 1.27 & 0.955 & 1625465 & 1625465 & 7.09 & 15.71\\
\hline
10.0 & $10^{-6}$ & $10^{-7}$ & 0.83 & 1.28 & 0.873 & 724918 & 7249184 & 7.41 & 15.72 \\
\hline
\end{tabular}}
\caption{Parameters for our simulations of decaying 3D MHD turbulence for the
three Prandtl numbers. These parameters are defined in the text.} 
\label{dec_para}
\end{table*}

It is useful to characterize the statistical properties of MHD turbulence
through the order-$p$, equal-time, structure functions of the longitudinal
component of increments of the velocity field, $\delta
u_{\parallel}(\bfx,\bfr,t) \equiv [\bfu(\bfx+\bfr,t)-\bfu(\bfx,t)].\bfr/r$, and
the magnetic field, $\delta b_{\parallel}(\bfx,\bfr,t)\equiv
[\bfb(\bfx+\bfr,t)-\bfb(\bfx,t)].\bfr/r$. These structure functions are
\begin{eqnarray}
{\cal S}^u_p(r) &\equiv& \la [\delta u_{\parallel}(\bfx,\bfr,t)]^p \ra \sim r^{\zeta^u_p};\\
\label{eq-ch5:stfnu}
{\cal S}^b_p(r) &\equiv& \la [\delta b_{\parallel}(\bfx,\bfr,t)]^p \ra \sim r^{\zeta^b_p};
\label{eq-ch5:stfnb}
\end{eqnarray}
they show power-law dependences on $r$ for the inertial range
$\eta_d^u,\eta_d^b \ll r \ll L$.  The angular brackets denote an average over
the statistically steady state of forced 3D MHD turbulence or an average over
different initial configurations for decaying 3D MHD turbulence.

In contrast with the scaling behaviors of correlation functions at conventional
critical points in equilibrium statistical mechanics, in 3D MHD turbulence the
structure functions ${\cal S}^u_p(r)$ and ${\cal S}^b_p(r)$ do not exhibit
simple scaling forms: numerical evidence suggests that these structure
functions exhibit multiscaling, with $\zeta^u_p$ and $\zeta^b_p$ nonlinear,
convex, monotonically increasing functions of $p$~\cite{sahoomhd,pouquet}. In
the absence of a mean magnetic field, an extension of the 1941 phenomenological
theory of Kolmogorov (K41)~\cite{k41} yields simple scaling with
$\zeta_p^{u,K41} = \zeta_p^{b,K41} = p/3$; but the measured values of
$\zeta_p^u$ and $\zeta_p^b$, e.g., deviate significantly from the K41
prediction, especially for $p > 3$ (see, e.g., Ref.~\cite{sahoomhd}).  Even
though the K41 phenomenology fails to capture the multiscaling of the
equal-time structure function ${\cal S}_p(r)$, it provides us with important
conceptual underpinnings for studies of homogeneous, isotropic 3D MHD
turbulence in the absence of a mean magnetic field. In the presence of a mean
magnetic field, Alfv\'en waves are present; they introduce a new time scale
into the problem, so Iroshnikov~\cite{irosh}, Kraichnan~\cite{kraichnan}, and
Dobrowlny, {\it et al.}~\cite{dob} have suggested that the energy spectrum
$E(k)$ can scale as $k^{-3/2}$, rather than $k^{-5/3}$, which follows from the
K41 phenomenology.  We refer the reader to Refs.~\cite{MHD-reviews,galtier} for
a discussion of the Dobrowlny-Iroshnikov-Kraichnan phenomenology.  We restrict
ourselves to the case when the mean magnetic field vanishes.

Several recent studies, especially on solar-wind data, have
yielded equal-time, two-point correlations that are consistent with the
phenomenological picture we have described above~\cite{space}. Shell
models, of the type we use here, have played a key role in understanding 
such correlations and emphasizing, e.g., the importance of the Hall
term in astrophysical MHD (see the elucidation of the
dual spectral scaling ranges, seen in solar-wind data, in the
careful shell-model investigations of 
Refs.~\cite{hallmhdshell1,hallmhdshell2,hallmhdshell3,hallmhdshell4}).
In the coming years, advances in probes for the solar wind and interplanetary
plasmas should lead to accurate measurements of not only spatial
correlations but also of spatiotemporal correlations of the turbulent fields in
these systems; it is in this light that our study, which characterizes
such spatiotemporal correlations in MHD turbulence and unearths their 
multiscaling nature, assumes special importance. 

\section{Dynamic Multiscaling\label{sec:dm}}

To set the stage for our discussions on time-dependent structure functions in
MHD turbulence, it is useful first to recall the results of analogous studies
in fluid turbulence. 

A na\"ive generalisation of K41 phenomenology to time-dependent velocity
structure functions in fluid turbulence yields simple dynamic scaling with a
dynamic exponent $z^{{\rm K41}}_p = 2/3$ for all orders $p$.  In order to
obtain non-trivial dynamic exponents, we need time-dependent velocity structure
functions and we must distinguish between Eulerian ($\cal E$), Lagrangian
($L$), and quasi-Lagrangian ($QL$) fields.  Eulerian fields yield trivial
dynamic scaling with $z^{\cal E} = 1$, for all $p$, because of the sweeping
effect and, hence, nontrivial dynamic multiscaling is possible only for
Lagrangian~\cite{book-pope} or quasi-Lagrangian~\cite{belinicher87} velocity
structure functions.  The latter are defined in terms of the quasi-Lagrangian
velocity ${\bf \hat u}$ that is related to its Eulerian counterpart $\bfu$ as
follows: 
\begin{equation}
{\hat \bfu}(\bfx,t) \equiv \bfu[\bfx + {\bf R}(t;\bfr_0,0),t] ,
\label{eq-ch5:qltrans}
\end{equation}
with ${\bf R}(t;{\bf r_0},0)$ the position at time $t$ of a Largrangian
particle that was at ${\bf r_0}$ at time $t = 0$.  Equal-time, quasi-Lagrangian
velocity structure functions are the same as their Eulerian
counterparts~\cite{rayprl,lvov93}.

For MHD turbulence we can define the order-$p$, time-dependent, structure
functions, for longitudinal, quasi-Lagrangian velocity and magnetic-field
increments as follows:
\begin{eqnarray}
{\mathcal F}^u_p(r,\{t_1,\ldots,t_p\}) \equiv
        \la [\delta \hat{u}_{\parallel}({\bf x},t_1,r) \ldots
              \delta \hat{u}_{\parallel}({\bf x},t_p,r)] \ra; 
\label{eq-ch5:Fp1}\\
{\mathcal F}^b_p(r,\{t_1,\ldots,t_p\}) \equiv
        \la [\delta \hat{b}_{\parallel}({\bf x},t_1,r) \ldots
              \delta \hat{b}_{\parallel}({\bf x},t_p,r)] \ra.
\label{eq-ch5:Fp}
\end{eqnarray}
(See Refs.~\cite{pandit08,ray08,rayprl} for similar structure functions for
fluid and passive-scalar turbulence.) In our studies of scaling properties, we
restrict $r$ to lie in the inertial range and consider, for simplicity, $t_1=t$ and
$t_2=\ldots=t_p=0$ and denote the structure functions, defined in
Eqs.~(\ref{eq-ch5:Fp1}) and (\ref{eq-ch5:Fp}),  by $F^u_p(r,t)$ and
$F^b_p(r,t)$, respectively. Given ${\mathcal F}^u_p(r,t)$ and ${\mathcal
F}^b_p(r,t)$, there are different ways of extracting time scales.  For example,
we can define order-$p$, degree-$M$ integral-time scale (superscript $I$) for
the velocity field as follows:
\begin{equation}
{\cal T}^{I,u}_{p,M}(r) \equiv
 \biggl[ \frac{1}{{\mathcal S}^u_p(r)}
\int_0^{\infty}{\mathcal F}^u_p(r,t)t^{(M-1)} dt \biggl]^{(1/M)};
\label{eq-ch5:timp-u}
\end{equation}
their analog for the magnetic field is
\begin{equation}
{\cal T}^{I,b}_{p,M}(r) \equiv
 \biggl[ \frac{1}{{\mathcal S}^b_p(r)}
\int_0^{\infty}{\mathcal F}^b_p(r,t)t^{(M-1)} dt \biggl]^{(1/M)}.
\label{eq-ch5:timp-b}
\end{equation}
If the integrals in Eqs.~(\ref{eq-ch5:timp-u}) and (\ref{eq-ch5:timp-b}) exist,
we can generalize the dynamic-scaling {\it Ansatz} (\ref{eq-ch5:ds}) at a
critical point to the following dynamic-multiscaling {\it Ans\"atze} for
homogeneous, isotropic MHD turbulence [for fluid turbulence
see Refs.~\cite{pandit08,ray08, rayprl}]. For the velocity integral-time scales we
assume
\begin{eqnarray}
{\cal T}^{I,u}_{p,M}(r) &\sim & r^{z^{I,u}_{p,M}};
\label{ziu} 
\end{eqnarray}
and, similarly, for the magnetic integral-time scales
\begin{eqnarray}
{\cal T}^{I,b}_{p,M}(r) &\sim & r^{z^{I,b}_{p,M}}.
\label{zib} 
\end{eqnarray}
These equations define, respectively, the integral-time dynamic-multiscaling 
exponents $z^{I,u}_{p,M}$ and  $z^{I,b}_{p,M}$. Time scales based on derivatives can
be defined as in Refs.~\cite{pandit08,ray08, rayprl}. For the purpose of 
illustrating dynamic multiscaling of structure functions in 3D MHD
turbulence, it suffices to use integral time scales, to which we restrict
ourselves in the remaining part of this paper. We return to the issue of 
other time-scales, e.g., the derivative time-scale in the concluding section of this 
paper.

\begin{figure*}
\begin{center}
\includegraphics[width=0.328\linewidth]{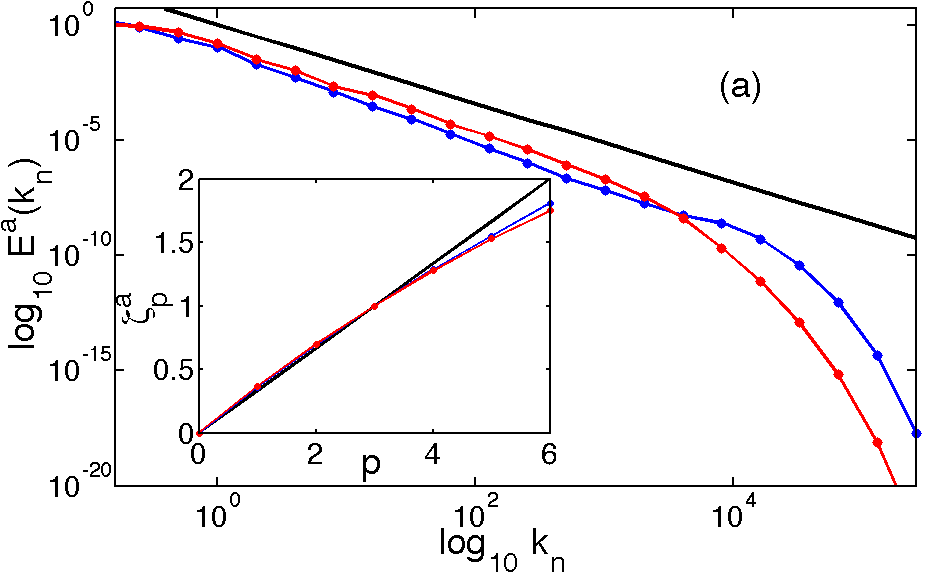}
\includegraphics[width=0.328\linewidth]{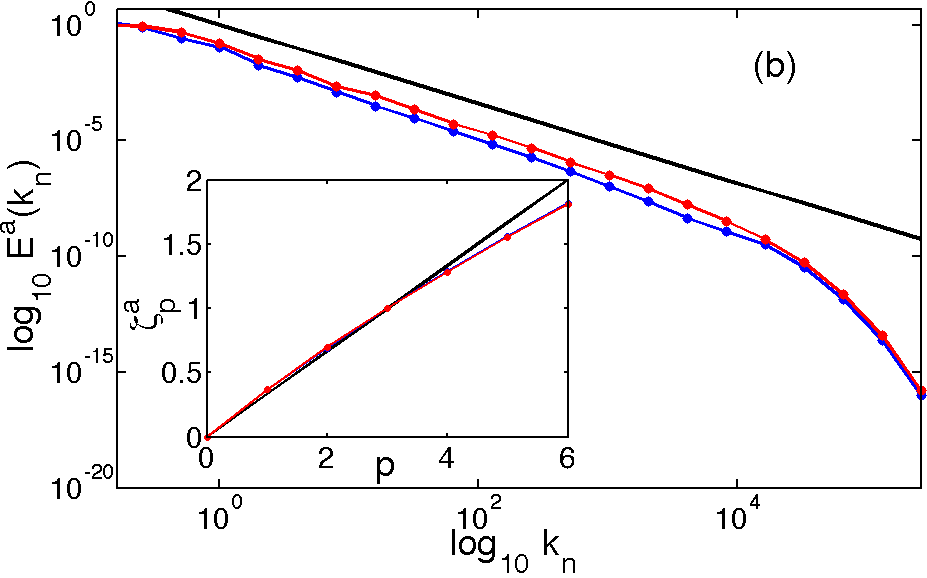}
\includegraphics[width=0.328\linewidth]{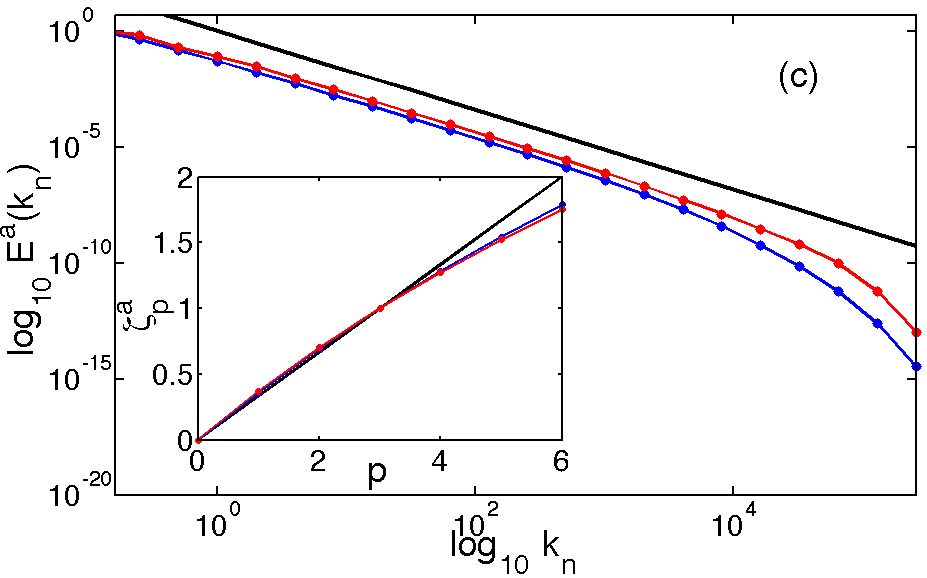}
\caption{(Color online) Log-log plots of the fluid kinetic (blue, filled circles) and
magnetic (red, filled squares) energy spectra versus the wave-vector for (a)
${\rm Pr_M}$ = 0.1, (b) ${\rm Pr_M}$ = 1.0, and (c) ${\rm Pr_M}$ = 10.0. The
black line is the K41 line that indicates $k^{-5/3}$ scaling. Insets : Plots of
the equal-time scaling-exponent ratios $\zeta_p^u/\zeta_3^u/$ (blue, filled
circles) and $\zeta_p^b/\zeta_3^b$ (red, filled squares), for the fluid and
magnetic equal-time structure functions, respectively, versus the order $p$,
for the same values of the Prandtl Number  ${\rm Pr_M}$; the black line denotes the K41
line $p/3$, which is shown for reference. Note: The negative values on the 
vertical axes are a result the logarithmic scales we use.} 
\label{dec_eps_spec}
\end{center}
\end{figure*}

\section{The 3D MHD shell model\label{sec:shell}}

We carry out extensive numerical simulations to obtain equal-time and dynamic
multiscaling exponents for a GOY-type ~\cite{book-frisch,goy1,goy2} shell model for
three-dimensional MHD turbulence~\cite{basu98}.  This shell model is defined on
a logarithmically discretized Fourier space labelled by scalar wave vectors
$k_n$ that are associated with the shells $n$. The dynamical variables are the
complex scalar shell velocities $u_n$ and magnetic fields $b_n$. The evolution
equations for $u_n$ and $b_n$ for 3D MHD are given by 
\begin{widetext}
\begin{eqnarray}
\frac{du_n}{dt}+\nu k_n^2u_n&=&\iota[A_n(u_{n+1}u_{n+2}-b_{n+1}b_{n+2})
+B_n(u_{n-1}u_{n+1}-b_{n-1}b_{n+1})
+C_n(u_{n-2}u_{n-1}-b_{n-2}b_{n-1})]^\ast ;\\
\label{eq:velshell}
\frac{db_n}{dt}+\eta k_n^2b_n&=&\iota[D_n(u_{n+1}b_{n+2}-b_{n+1}u_{n+2})
+ E_n(u_{n-1}b_{n+1}-b_{n-1}u_{n+1})
+F_n(u_{n-2}b_{n-1}-b_{n-2}u_{n-1})]^\ast ;
\label{eq-ch5:magshell}
\end{eqnarray}
\end{widetext}
here $k_n = k_0 2^n\/$, $k_0$ = 1/16, complex conjugation is 
denoted by $\ast$, and the coefficients
\begin{widetext}
\begin{equation}
A_n=k_n,\;\;B_n=-k_{n-1}/2,\;\;C_n=-k_{n-2}/2,\;\;
D_n=k_n/6,\;\;E_n=k_{n-1}/3,\;\;F_n=-2k_{n-2}/3,
\end{equation}
\end{widetext}
are chosen to conserve the shell-model analogs of the total energy
$E_T=E_u+E_b\equiv(1/2)\sum_n (|u_n|^2 + |b_n|^2)$, cross helicity
$H_C\equiv(1/2)\sum_n(u_nb_n^*+u_n^*b_n)$, and magnetic helicity
$H_M\equiv\sum_n(-1)^n|b_n|^2/k_n$, in the inviscid, unforced limit, and $1\leq
n \leq N$.  The logarithmic discretization of Fourier space allows us to reach
very high Reynolds numbers in numerical simulations of this shell model even
with $N = 22$.

\begin{figure}
\begin{center}
\includegraphics[width=1.0\columnwidth]{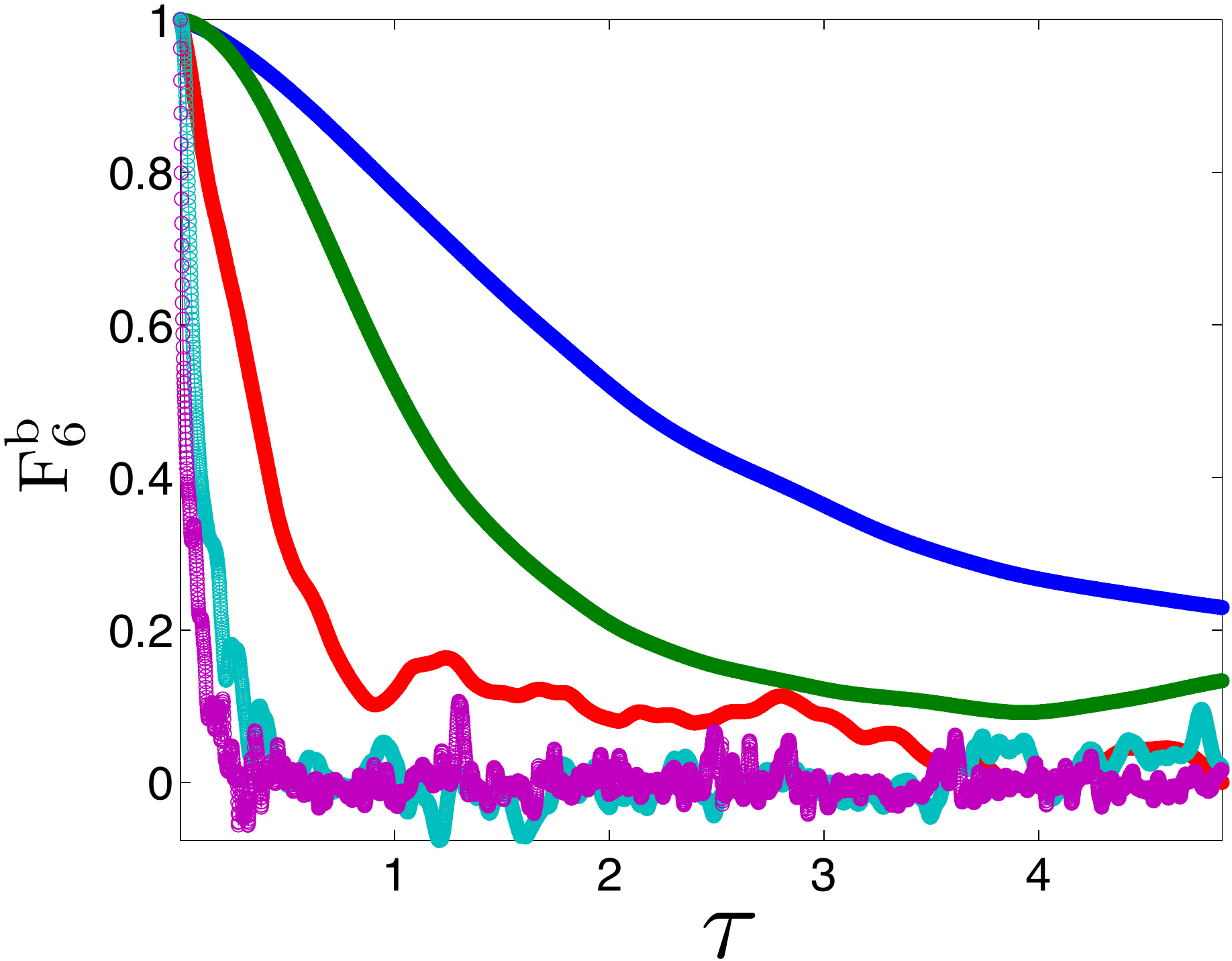}
\caption{(Color online) A representative plot of the normalised, time-dependent 
sixth-order structure function $F_6^b$, for the magnetic field, versus the normalised time $\tau = t/\tau_b$, where 
$\tau_b$ is the characteristic small time-scale associated with the magnetic field. 
The different curves correspond to shell numbers 5 (uppermost, in blue), 6, 9, 12, and 15 (lowermost, in magenta). The 
plot is shown for ${\rm Pr_M}$ = 1.0.} 
\label{Q6}
\end{center}
\end{figure}

The shell-model equations allow for direct interactions only between nearest-
and next-nearest-neighbor shells. By contrast, in the Fourier transform of the
3D MHD equations, every Fourier mode of the velocity and magnetic fields is
coupled to every other Fourier mode directly. This leads to direct sweeping of
small eddies by large ones and to the trivial dynamic scaling of Eulerian
velocity structure functions that we have mentioned above. Our shell model does
not have direct sweeping in this sense; thus, such shell models are sometimes
thought of as simplified, quasi-Lagrangian versions of their parent
hydrodynamic equations. Therefore, we expect nontrivial dynamic multiscaling
for structure functions in this MHD shell model as has been
found~\cite{pandit08,ray08} for the GOY model for fluid turbulence.

\begin{figure*}
\begin{center}
\includegraphics[width=0.329\linewidth]{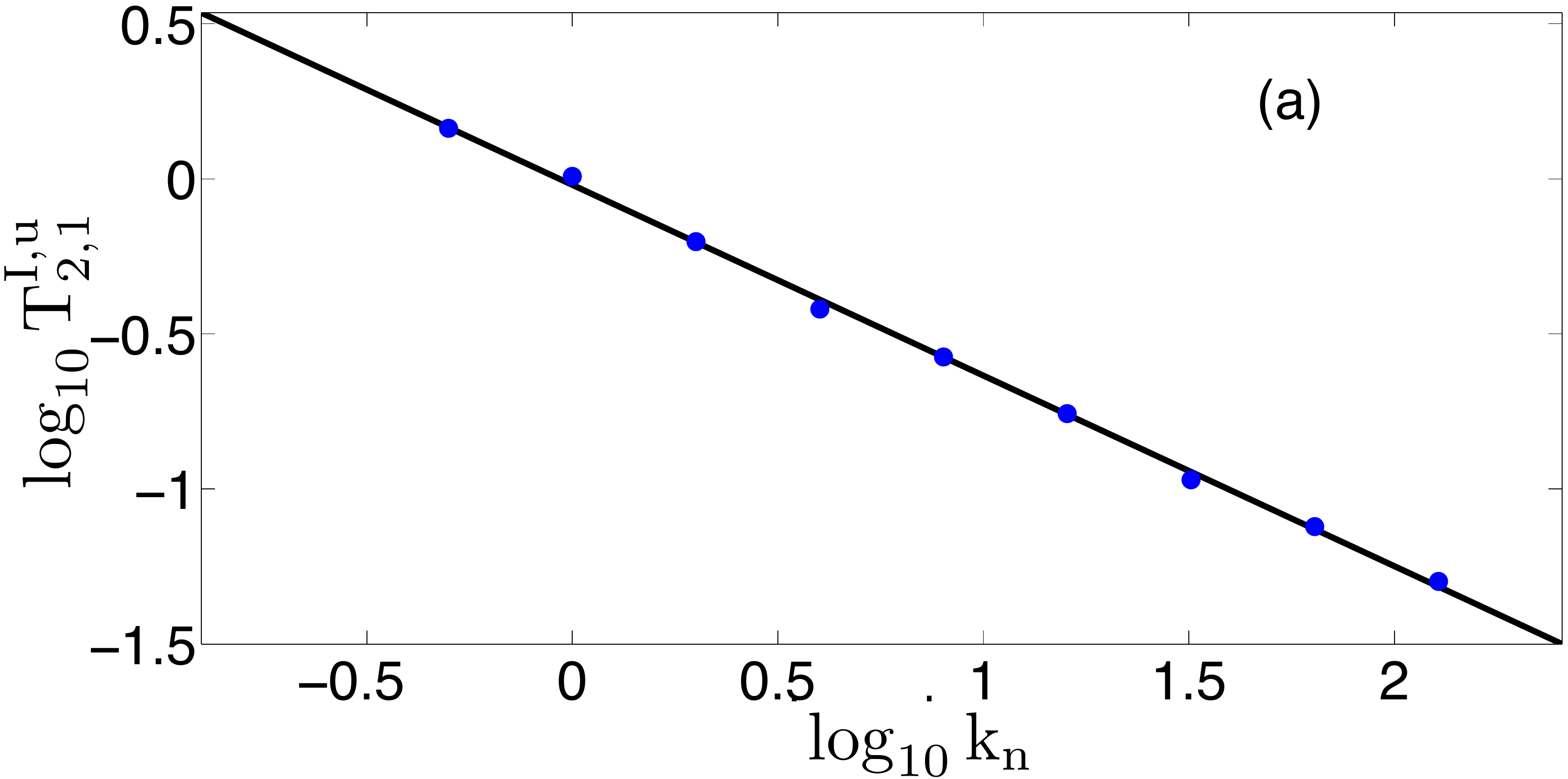}
\includegraphics[width=0.329\linewidth]{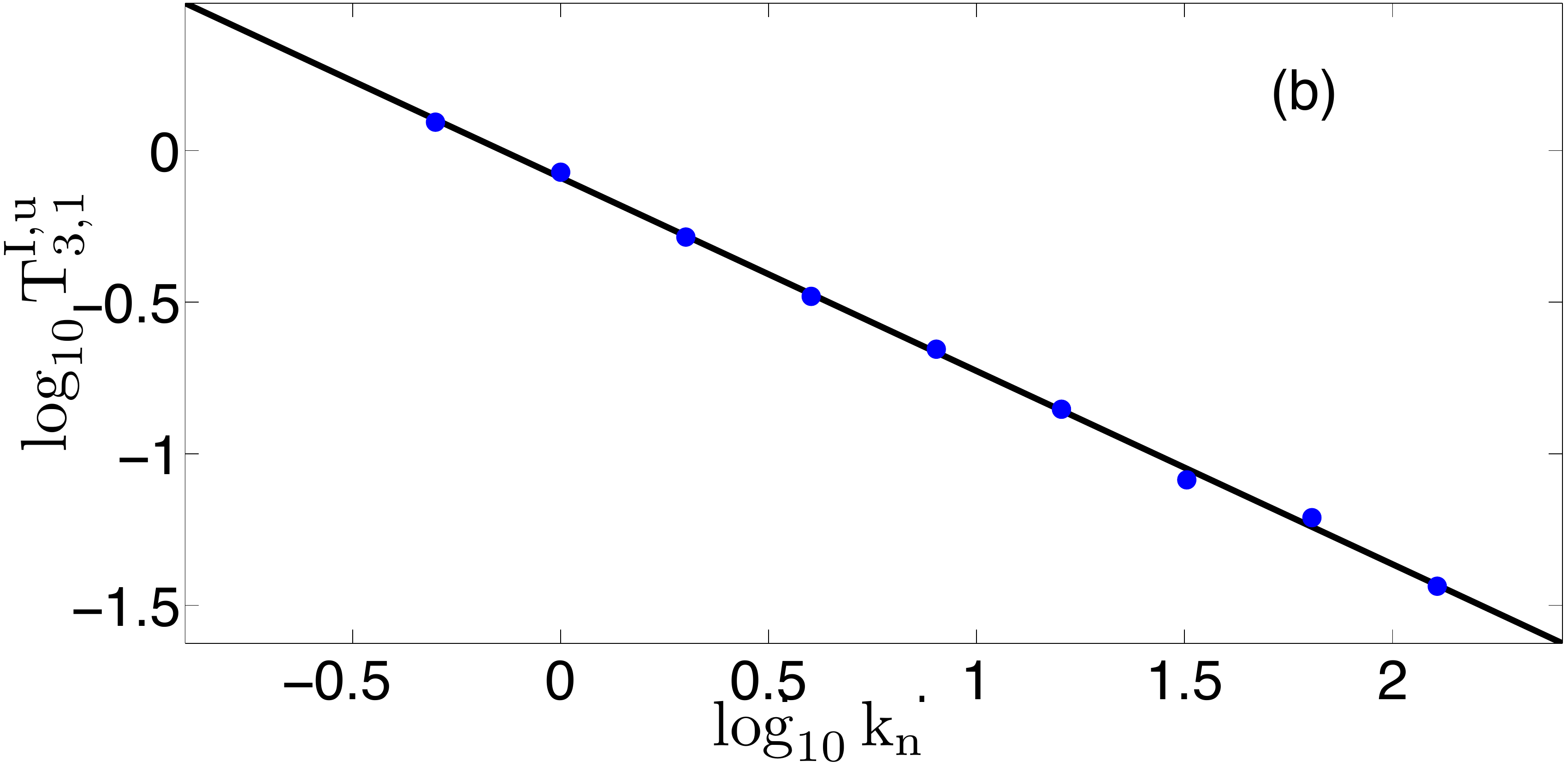}
\includegraphics[width=0.329\linewidth]{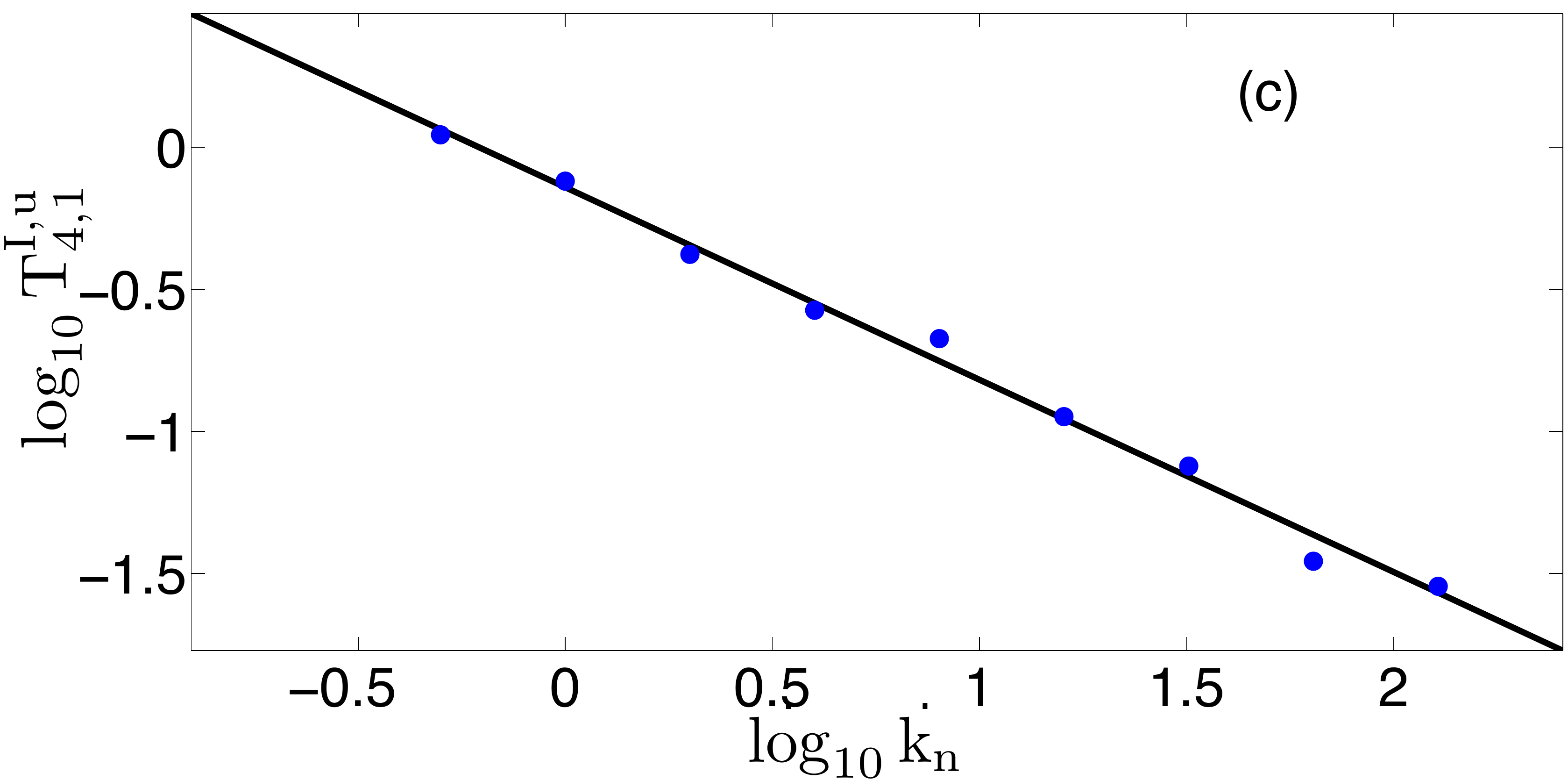}
\caption{(Color online) Log-log plots of the fluid  integral-time scale $T^{I,u}_{p,1}$ versus
the wave-vector for (a) ${\rm Pr_M}$ = 0.1 and $p$ = 2, (b) ${\rm Pr_M}$ = 1.0
and $p$ = 3,  and (c) ${\rm Pr_M}$ = 10.0 and $p$ = 4.} \label{dec_TIu}
\end{center}
\end{figure*}

The equal-time, velocity and magnetic field structure functions,  in the MHD
shell model, are defined as  
\begin{equation}
S^u_p(k_n) \equiv \la [u_n(t)u^{\ast}_n(t)]^{p/2} \ra
\label{eq-ch5:eqsfu}
\end{equation}
and
\begin{equation}
S^b_p(k_n) \equiv \la [b_n(t)b^{\ast}_n(t)]^{p/2} \ra,
\label{eq-ch5:eqsfb}
\end{equation}
respectively.  For shells lying in the inertial range, we obtain the equal-time
scaling exponents via   $S^u_p(k_n) \sim k_n^{-\zeta_p^u}$ and $S^b_p(k_n) \sim
k_n^{-\zeta_p^b}$.  The three cycles~\cite{kadanoff95} in the static solutions
of shell models lead to rough, period-three oscillations in $S^u_p(k_n)$ and
$S^b_p(k_n)$. Hence, in order to obtain scaling regions without such 
oscillations, we generalize the
modified structure functions suggested in Ref.~\cite{kadanoff95} for the GOY
model for our MHD shell model. In particular, we use
\begin{eqnarray}
\Sigma^u_p &\equiv&  \la|{\Im}[u_{n+2}u_{n+1}u_n - (1/4)u_{n-1}u_nu_{n+1}]|^{p/3}\ra,  
\nonumber \\
\Sigma^b_p &\equiv&  \la|{\Im}[b_{n+2}b_{n+1}b_n - (1/4)b_{n-1}b_nb_{n+1}]|^{p/3}\ra; 
\label{eq-ch5:sigmab}
\end{eqnarray}
in these structure functions, the period-three oscillations are effectively
filtered out. We can also obtain the exponent ratios $\zeta^u_p/\zeta^u_3$ and
$\zeta^b_p/\zeta^b_3$ by using the extended-self-similarity (ESS) procedure in
which we plot, respectively, $S_p^u$ versus $S_3^u$ and $S_p^b$ versus $S_3^b$
(see, e.g., Refs.~\cite{sahoomhd,ess}).

We then employ the order-$p$, time-dependent structure functions for the 3D MHD
shell model, namely,  
\begin{eqnarray}
F^u_p(k_n,t_0,t) &\equiv& \la [u_n(t_0)u^{\ast}_n(t_0 + t)]^{p/2} \ra; 
\label{eq-ch5:stfn-tdu} \\
F^b_p(k_n,t_0,t) &\equiv& \la [b_n(t_0)b^{\ast}_n(t_0 + t)]^{p/2} \ra,
\label{eq-ch5:stfn-td}
\end{eqnarray}
to obtain the integral time scales.
In general, these time-dependent structure functions are complex; we find that their imaginary parts 
are much smaller than their real parts; therefore, in all our studies, we restrict ourselves to 
the real parts of these time-dependent structure functions.
References~\cite{mitra04,mitra05,pandit08,ray08, rayprl}  have used such
time-dependent structure functions to verify bridge relations for fluid and
passive-scalar turbulence. Both equal-time and time-dependent structure
functions can also be obtained for the shell-model analogs of the Els\"asser
variables $z^{\pm}_n = u_n \pm b_n$.

The root-mean-square velocity and the root-mean-square magnetic field are
defined, respectively, as $u_{\rm rms} = [\la \sum_n |u_n|^2 \ra]^{1/2}$ and
$b_{\rm rms} = [\la \sum_n |b_n|^2 \ra]^{1/2}$. The Taylor microscale $\lambda
\equiv \frac{\la\sum_n(|u_n^2|/k_n^2)\ra}{\la\sum_n(|u_n^2|/k_n)\ra}$, 
and the magnetic and fluid Reynolds-number, based on the Taylor microscale are,
respectively, ${\rm Re}_{\lambda,u} = u_{\rm rms} \lambda/\nu$ and  ${\rm
Re}_{\lambda,b} = b_{\rm rms} \lambda/\eta$.  Finally, we define the
characteristic time scales for the fluid and the magnetic field via $\tau_u
\equiv (u_{\rm rms}k_1)^{-1}$ and $\tau_b \equiv (b_{\rm rms}k_1)^{-1}$,
respectively. The angular brackets in the above definitions imply an average
over different initial configurations because we study decaying MHD turbulence
in the remaining part of this paper.

\begin{table}

{\begin{tabular}{|c|c|c|c|c|}
\hline
$p$ & $\zeta^u_p$ (ESS) & $\zeta^b_p$ (ESS) & $z^{I,u}_{p,1}$ & $z^{I,b}_{p,1}$ \\
\hline
1 & 0.362 $\pm$ 0.005 & 0.367 $\pm$ 0.004 & 0.593 $\pm$ 0.001 & 0.597 $\pm$ 0.001\\
\hline
2 & 0.694 $\pm$ 0.004 & 0.698 $\pm$ 0.004 & 0.622 $\pm$ 0.004 & 0.632 $\pm$ 0.002 \\
\hline
3 & 1.0000 & 1.0000 & 0.623 $\pm$ 0.007 & 0.629 $\pm$ 0.003\\
\hline
4 & 1.287 $\pm$ 0.008 & 1.277$\pm$ 0.006 & 0.62 $\pm$ 0.01 & 0.627 $\pm$ 0.007\\
\hline
5 & 1.548 $\pm$ 0.009 & 1.528 $\pm$ 0.009 & 0.62 $\pm$ 0.01 & 0.629 $\pm$ 0.009 \\
\hline
6 & 1.81 $\pm$ 0.03 & 1.75 $\pm$ 0.03 & 0.61 $\pm$ 0.02 & 0.63 $\pm$ 0.01 \\ 
\hline
\end{tabular}}
\caption{The equal-time (ESS) and the integral-time dynamic multiscaling exponents 
for the fluid and the magnetic field from our simulations of decaying MHD turbulence for 
the magnetic Prandtl Number ${\rm Pr_M = 0.1}$.}
\label{dec_expo0}
\end{table}

\section{Results\label{sec:results}}

\begin{table}
{\begin{tabular}{|c|c|c|c|c|}
\hline
 $p$ & $\zeta^u_p$ (ESS) & $\zeta^b_p$ (ESS) & $z^{I,u}_{p,1}$ & $z^{I,b}_{p,1}$ \\
\hline
1 & 0.361 $\pm$ 0.003 & 0.367 $\pm$ 0.003 & 0.616 $\pm$ 0.005 & 0.618 $\pm$ 0.002 \\
\hline
2 & 0.692 $\pm$ 0.004 & 0.696 $\pm$ 0.003 & 0.641 $\pm$ 0.003 & 0.646 $\pm$ 0.002 \\
\hline
3 & 1.0000 & 1.0000 & 0.645 $\pm$ 0.004 & 0.641 $\pm$ 0.002  \\
\hline
4 &  1.288 $\pm$ 0.007 & 1.284 $\pm$ 0.006  & 0.643 $\pm$ 0.006 & 0.634 $\pm$ 0.003  \\
\hline
5 & 1.564 $\pm$ 0.009  & 1.554 $\pm$ 0.009 & 0.639 $\pm$ 0.007 & 0.626 $\pm$ 0.005  \\
\hline
6 & 1.82 $\pm$ 0.01 & 1.81 $\pm$ 0.01 & 0.633 $\pm$ 0.009 & 0.617 $\pm$ 0.009 \\ 
\hline
\end{tabular}}
\caption{The equal-time (ESS) and the integral-time dynamic multiscaling
exponents for the fluid and the magnetic field from our simulations of decaying MHD
turbulence for the magnetic Prandtl Number ${\rm Pr_M = 1.0}$.}
\label{dec_expo1}
\end{table}

We consider an unforced 3D MHD shell-model (with 22 shells) for 3D MHD  with
the initial shell-model velocities and magnetic fields $u_n = k_n^{-1/3} \exp(i
\varphi_n)$ and $b_n = k_n^{-1/3} \exp(i \vartheta_n)$, with $\varphi_n$ and
$\vartheta_n$ random phases distributed uniformly on the interval $[0,2\pi)$,
respectively.  We choose a time step $\delta t = 10^{-4}$.  We obtain results
for ${\rm Pr_M}$ equal to 0.1, 1.0, and 10.0. The various parameters of these
simulations are given in Table \ref{dec_para}. 

\begin{figure*}
\begin{center}
\includegraphics[width=0.329\linewidth]{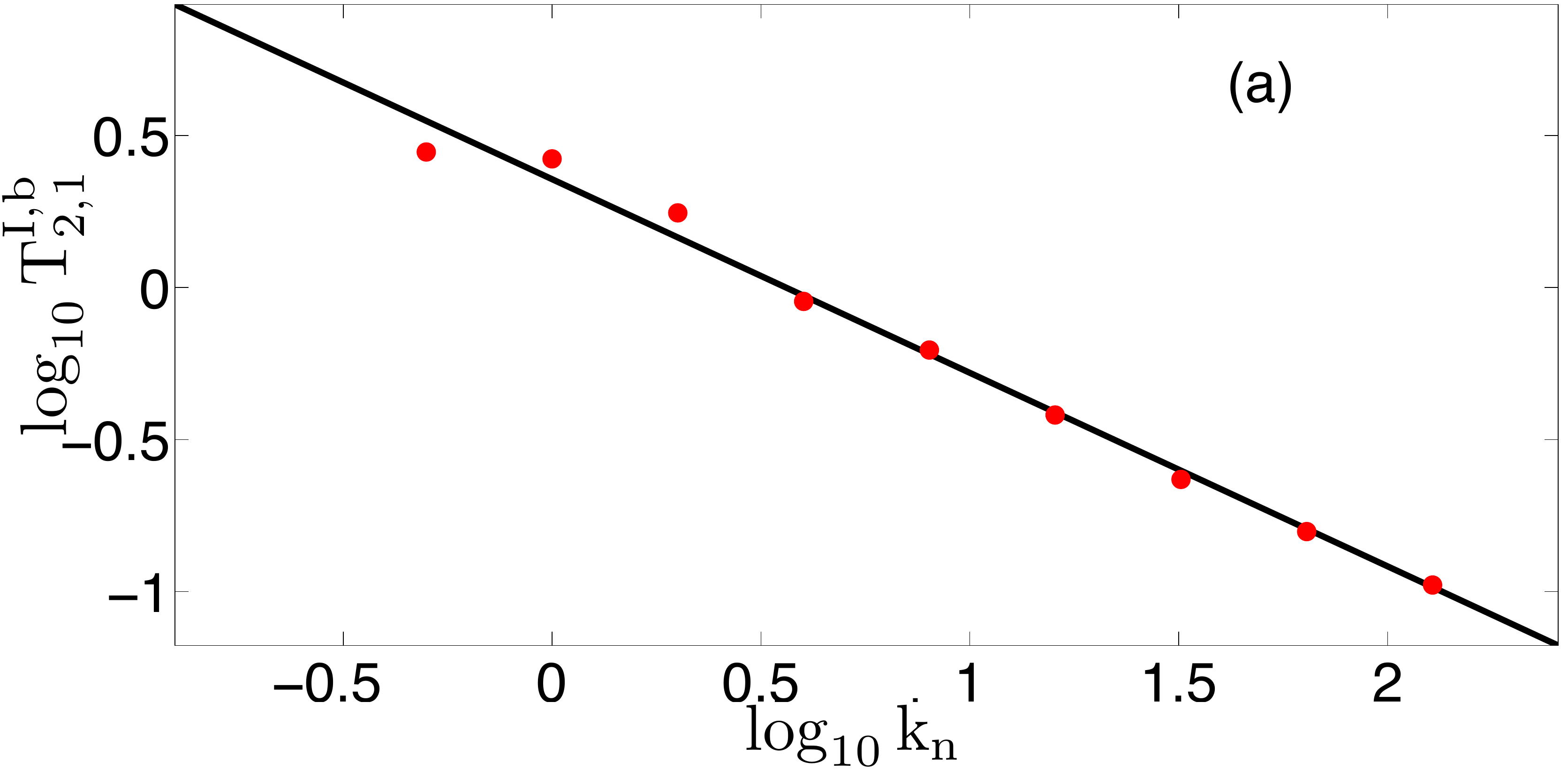}
\includegraphics[width=0.329\linewidth]{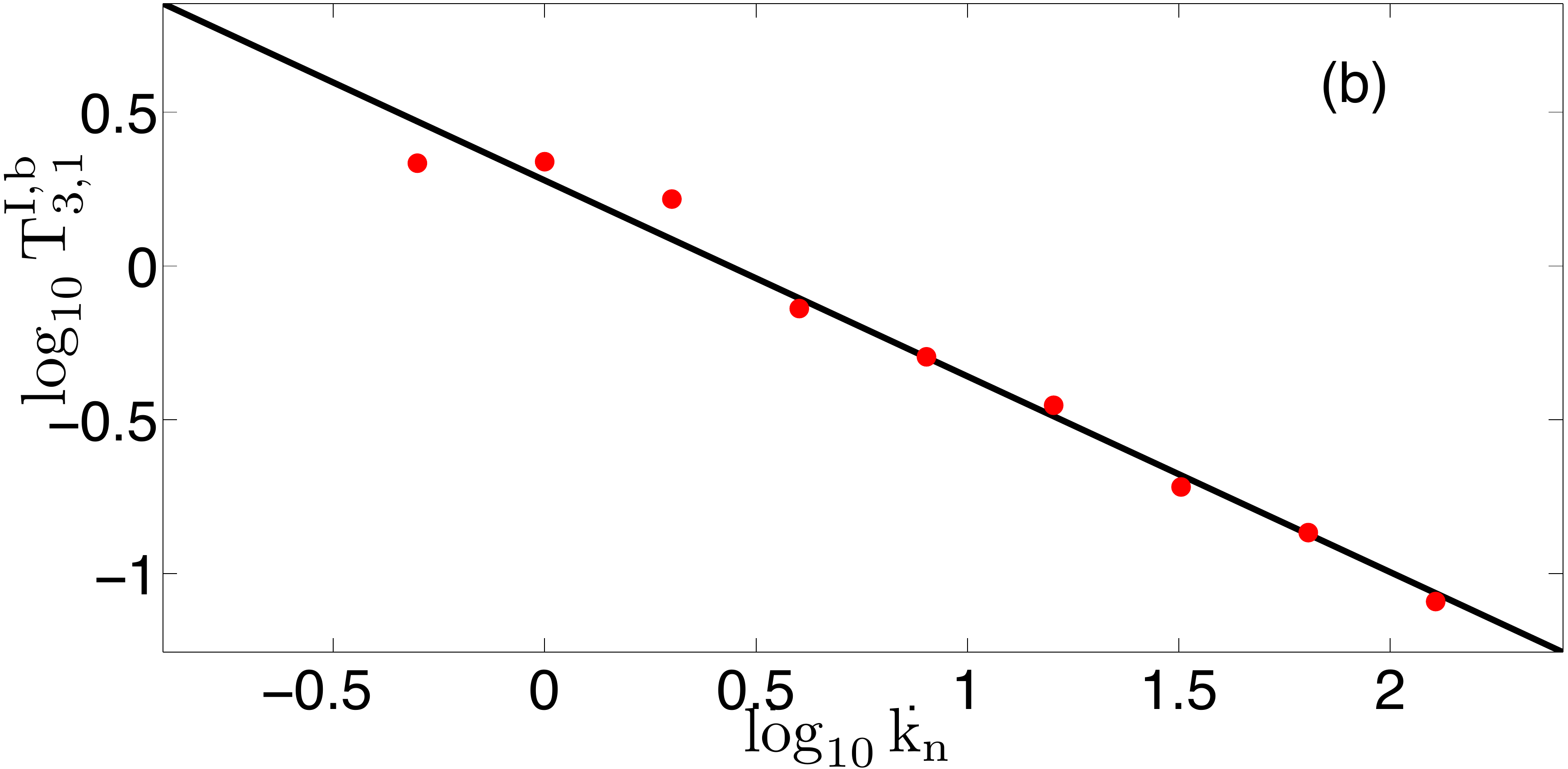}
\includegraphics[width=0.329\linewidth]{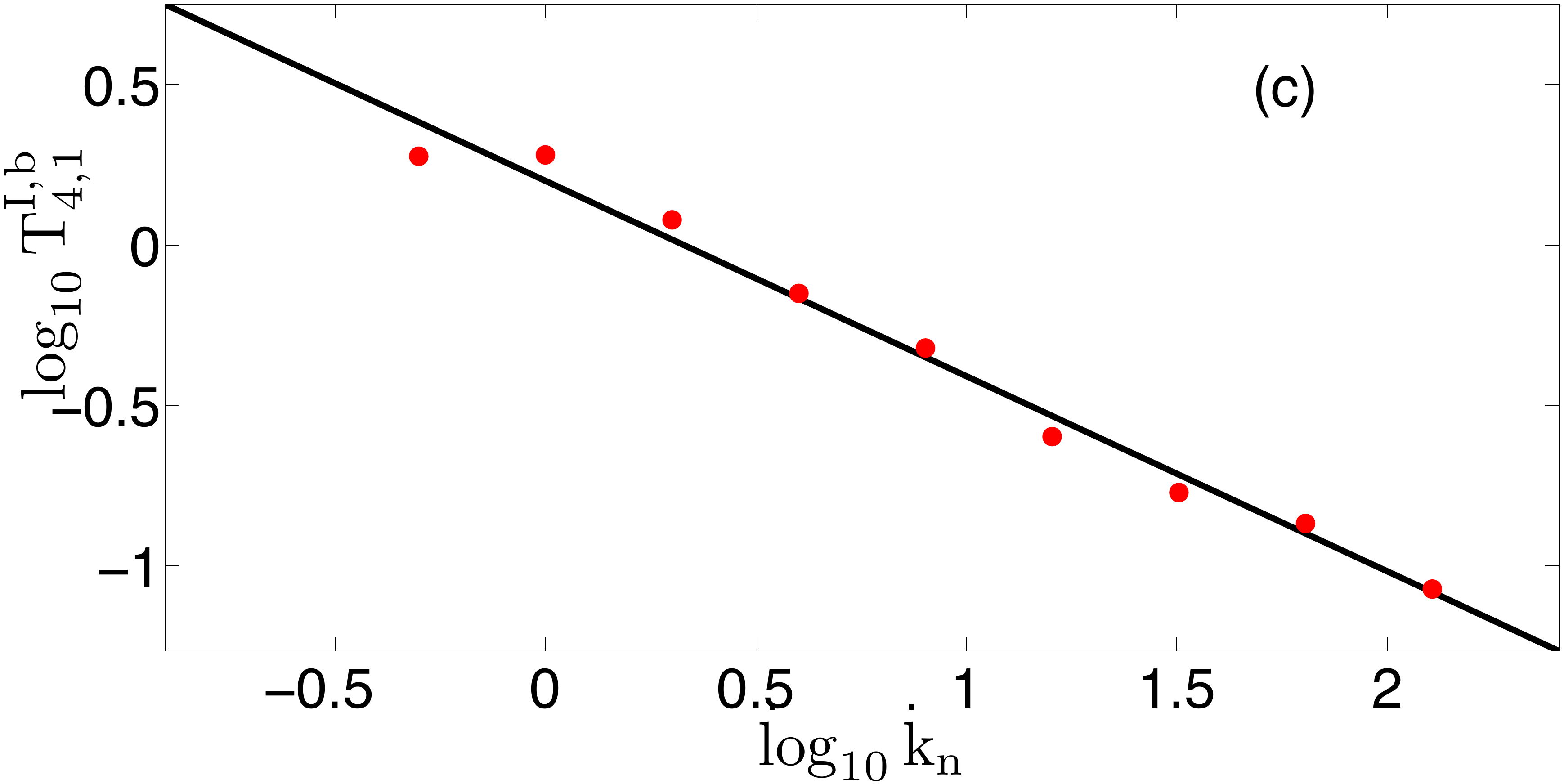}
\caption{(color online) Log-log plots of the fluid  integral-time scale $T^{I,u}_{p,1}$ versus
the wave-vector for (a) ${\rm Pr_M}$ = 0.1 and $p$ = 2, (b) ${\rm Pr_M}$ = 1.0
and $p$ = 3,  and (c) ${\rm Pr_M}$ = 10.0 and $p$ = 4.} \label{dec_TIb}
\end{center}
\end{figure*}

We calculate statistical quantities, like the energy spectrum, the equal-time structure
functions, the time-dependent structure functions, and the various
exponents extracted from them, only after the magnetic and fluid
kinetic energy dissipation rates have reached their peaks.  It is important to
note that the peak in the dissipation rates, for both the fluid kinetic energy
and the magnetic energy, occur at the same time here.  

This time, at which the peak occurs, signals cascade completion, i.e., after this time,
inertial-range fluid kinetic and the magnetic-energy  spectra 
display K41 scaling with intermittency corrections. In
Fig.~(\ref{dec_eps_spec}) we show representative plots of the fluid and
magnetic energy spectra for different values of ${\rm Pr_M}$. 
We note that 
our spectral scaling exponents are compatible with the K41 result plus 
a small intermittency correction.  The equal-time
exponent-ratios, $\zeta_p^u/\zeta_3^u$ and $\zeta_p^b/\zeta_3^b$, which we obtain by
using the ESS procedure and from Eqs.~(\ref{eq-ch5:sigmab}), are shown in the insets of
Fig.~(\ref{dec_eps_spec}), for the three values of ${\rm Pr_M}$ used in our
simulations; we list these ratios in Tables~\ref{dec_expo0}, \ref{dec_expo1},
and \ref{dec_expo2}; they are in agreement with previous studies
(see, e.g.,Ref.~\cite{sahoomhd} and references therein).

\begin{table}
{\begin{tabular}{|c|c|c|c|c|}
\hline
$p$ & $\zeta^u_p$ (ESS) & $\zeta^b_p$ (ESS) & $z^{I,u}_{p,1}$ & $z^{I,b}_{p,1}$ \\
\hline
1 & 0.356 $\pm$ 0.005 & 0.369 $\pm$ 0.005 & 0.653 $\pm$ 0.003 & 0.633 $\pm$  0.002 \\
\hline
2 & 0.690 $\pm$ 0.006 & 0.700 $\pm$ 0.006 & 0.673 $\pm$ 0.004 & 0.657 $\pm$ 0.002  \\
\hline
3&  1.0000 & 1.0000 & 0.679 $\pm$ 0.005 & 0.658 $\pm$ 0.004 \\
\hline
4 &  1.284 $\pm$ 0.008 & 1.272 $\pm$ 0.008 & 0.687 $\pm$ 0.009  & 0.660 $\pm$ 0.006  \\
\hline
5 & 1.543 $\pm$ 0.009 & 1.521 $\pm$ 0.009 & 0.70 $\pm$ 0.02 & 0.662 $\pm$ 0.009 \\
\hline
6 &  1.79 $\pm$ 0.02 & 1.75 $\pm$ 0.03  & 0.70 $\pm$ 0.02 & 0.66 $\pm$ 0.01 \\
\hline
\end{tabular}}
\caption{The equal-time (ESS) and the integral-time dynamic multiscaling
exponents for the fluid and the magnetic field from simulations of decaying MHD
turbulence for Prandtl Number  ${\rm Pr_M = 10}$.}
\label{dec_expo2}
\end{table}

We now turn our attention to the time-dependent structure functions (representative 
plots of which for the magnetic field and different shell numbers for order-6 are shown in Fig.~\ref{Q6}), whence we
calculate the integral-time scales (cf.
Eqs.~(\ref{eq-ch5:timp-u}-\ref{eq-ch5:timp-b}) with $M = 1$)
\begin{equation}
T^{I,u}_{p,1}(k_n) = \int_0^{t_\mu}F^u_p(k_n,t)dt;
\label{eq-ch5:intscaleu}
\end{equation}
and
\begin{equation}
T^{I,b}_{p,1}(k_n) = \int_0^{t_\mu}F^b_p(k_n,t)dt; 
\label{eq-ch5:intscaleb}
\end{equation}
we drop the index $M$ for notational convenience; henceforth we use $M =
1$.  In the above definitions, $t_{\mu}$ is the time at which $F^u_p(n,t) =
\mu$ (or $F^b_p(n,t) = \mu$) with $0\leq \mu \leq 1$. In principle we should
use $\mu = 0$, i.e., $t_{\mu} = \infty$, but this is not possible in any
numerical calculation because $F^u_p$ and $F^b_p$ cannot be obtained accurately
for large $t$. We use $\mu = 0.7$; and we have checked in representative cases
that our results do not change for $0.65 < \mu < 0.8$.  Representative log-log
plots of such integral-time scales  are shown for various orders $p$ and ${\rm Pr_M}$
for the fluid in Fig.~(\ref{dec_TIu}) and for the magnetic field in
Fig.~(\ref{dec_TIb}). We see rather clean inertial-range scaling for both these
quantities; and from numerical fits we obtain the dynamic multiscaling exponents
$z^{I,u}_{p,1}$ and $z^{I,b}_{p,1}$. The values of all these exponents, and
their dependence on ${\rm Pr_M}$, are shown in Tables~\ref{dec_expo0},
\ref{dec_expo1}, and \ref{dec_expo2}.  The mean values of these exponents are
obtained from $50$ different sets of statistically independent data; the mean of
these is quoted as the exponent and their standard deviation as the error bar.

If we consider fluid turbulence alone, the analogs of these
dynamic-multiscaling exponents satisfy linear bridge relations obtained in
Ref.~\cite{mitra04,pandit08,ray08}, which follow from a
generalization~\cite{lvov97,mitra04,pandit08,ray08} of the multifractal
formalism~\cite{book-frisch} for turbulence.  Unfortunately, this way of
obtaining bridge relations cannot be generalized to the case of MHD turbulence:
We must now deal with a joint multifractal distribution of both the velocity
and the magnetic-field variables. To obtain the analogs of the fluid-turbulence
bridge relations given above, additional decoupling approximations must be made
for averages of products of powers of velocity and magnetic-field increments;
such approximations have been used to obtain bridge relations for the case of
passive-scalar turbulence~\cite{mitra05,ray08}.  However, it is not possible to
justify such additional decoupling approximations for the case of MHD
turbulence because the magnetic field is actively advected by the fluid and, in
turn, affects the advecting velocity field.

\begin{figure}
\begin{center}
\includegraphics[width=1.0\columnwidth]{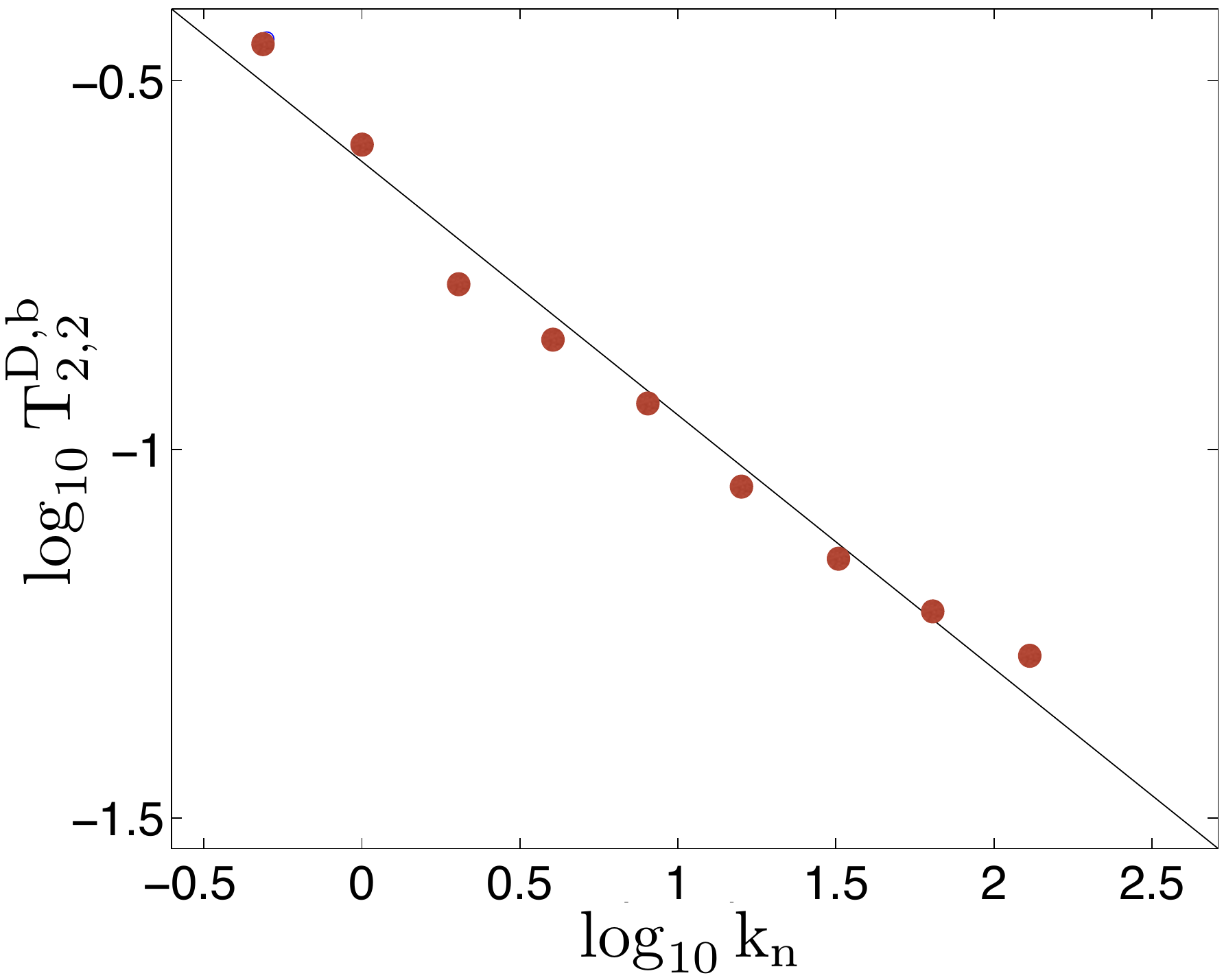}
\caption{(color online) Log-log plot of the magnetic derivative-time scale $T^{D,b}_{2,1}$ versus
the wave-vector for ${\rm Pr_M}$ = 1.0. The thick black line indicates a slope of 
$z^{D,b}_{2,2} = 0.35$.}
\label{Db2}
\end{center}
\end{figure}

\section{Conclusions}\label{sec:concl}

Spatiotemporal correlation or structure functions are
important measures by which we probe the statistical properties of turbulent
flows, which arise from the dynamics and the coupling of the multiple spatial
and temporal scales. We have developed the formalism for exploring the dynamic
multiscaling of time-dependent structure functions that characterize MHD
turbulence. Such dynamic multiscaling has been studied hitherto only for fluid
and passive-scalar turbulence. Furthermore, we have carried out extensive
numerical studies of such time-dependent structure functions in a shell model
for 3D MHD turbulence ~\cite{basu98} and thus shown that an infinity of
dynamic-multiscaling exponents is required to characterize the multiscaling of
time-dependent structure functions. Finally, we have demonstrated that these
exponents depend on the precise way in which time scales are extracted from
these structure functions. 

The measurement of two-point, spatiotemporal fluctuations, via
time-dependent structure functions and their associated scaling
exponents~\cite{rayprl,calzavarini,ann-rev}, is a challenging numerical task,
especially if quasi-Lagrangian structure functions have to be obtained from
direct numerical simulation (DNS) of hydrodynamical equations, as has been done
for fluid turbulence in two and three dimensions~\cite{rayprl,calzavarini}.
Given current computational resources, it is best to begin with shell-model
studies, which have very special advantages compared to quasi-Lagrangian DNSs.
Apart from the enormous range in Reynolds number that we can cover in shell
models, the basic quasi-Lagrangian structure of shell models helps to eliminate
trivial dynamic scaling, which arises because of the sweeping effect that we
have discussed above.  

Given recent advances in observational techniques for space
plasmas and the solar winds~\cite{space}, we expect that direct measurements of
time-dependent structure functions should be possible in these systems.  Our
study of these structure functions in MHD turbulence is directly relevant to
such experiments. It is important to realize the importance of such a
calculation within the framework of shell models. For example, a recent
shell-model study of Hall-MHD turbulence has provided an explanation of the
spectral properties of turbulence in the solar wind~\cite{hallmhdshell4}.

Our numerical simulations of the MHD-shell-model equations are several orders
of magnitude longer than those in earlier studies~\cite{basu98} and our
Reynolds numbers are much higher than those achieved there. Given that the MHD
system is considerably more complicated than its Navier-Stokes counterpart, it
is not surprising that it is much harder to characterize dynamic multiscaling
in the former than it is in the latter from \textit{both theoretical and
numerical points of view}.  As we have mentioned above, we cannot obtain for
MHD turbulence the analogs of bridge relations~\cite{mitra04,pandit08,ray08},
which relate equal-time and dynamic-multiscaling exponents in fluid and
passive-scalar turbulence.  The exploration of dynamic multiscaling for the
full MHD equations, via quasi-Lagrangian approaches~\cite{rayprl,calzavarini},
remains a challenge.

In the case of previous shell-model studies of fluid and passive-scalar 
turbulence, it has been possible to extract, additionally the derivative-time-scale and the associated 
scaling exponents for the same~~\cite{mitra04,mitra05,pandit08,ray08}. We find similar derivative time-scale 
exponents, defined via $T^{D,b}_{p,2}(k_n) =   \left[ \frac{1}{{\mathcal S}^{b}_p(k_n)}
\frac{\partial^2 {\mathcal F}^{b}_p(k_n,t)}{\partial t^2}
\biggl {|}_{t=0} \right]^{-1/2}$  and 
$T^{D,u}_{p,2}(k_n) =  \left[ \frac{1}{{\mathcal S}^{u}_p(k_n)}
\frac{\partial^2 {\mathcal F}^{u}_p(k_n,t)}{\partial t^2}
\biggl {|}_{t=0} \right]^{-1/2}$, for our MHD shell model. 
In Fig.~\ref{Db2} we show a representative log-log plot of $T^{D,b}_{p,2}$ versus the wavenumber $k_n$ to yield the 
scaling exponent $z^{D,b}_{p,2}$. We will, in future work, address systematically the issue of the derivative time-scale of different 
orders and explore the hierarchy of exponents for the integral and derivative time-scales.

SSR and RP acknowledge the support of the Indo-French Center for Applied
Mathematics (IFCAM); RP thanks the Department of Science and Technology, the
Council of Scientific and Industrial Research (India) for support and SERC
(IISc) for computational resources; and SSR thanks the AIRBUS Group Corporate
Foundation Chair in Mathematics of Complex Systems established in the ICTS as
well as DST (India) project ECR/2015/000361 for support. GS acknowledges
funding from the European Research Council under the European Union's Seventh
Framework Programme, ERC Grant Agreement No 339032. 


\end{document}